\documentclass[aps,twocolumn,showpacs,nofootinbib]{revtex4-1}

\usepackage{color} 
\newcommand{\blue}[1]{{\color{blue} #1}}
\definecolor{green}{rgb}{0,.5,0}

\definecolor{red}{rgb}{1,0,0}

\usepackage{graphicx}
\usepackage[subfigure]{graphfig}
\usepackage{epsfig}
\usepackage{dcolumn}
\usepackage{amsmath}

\newcommand{\dslash}[1]{{#1\!\!\!/}}
\def\bea{\begin{eqnarray}}

\def\eea{\end{eqnarray}}

\immediate\write18{texcount -inc -incbib 
-sum borra.tex > /tmp/wordcount.tex}

\begin{document}

\title{\vspace{1.0in}The 1-loop correction of the QCD energy momentum tensor with the overlap fermion and HYP smeared Iwasaki gluon}

\author{{Yi-Bo Yang$^{1}$, Michael Glatzmaier$^{1}$, Keh-Fei Liu$^{1}$, Yong Zhao$^{2,3,4}$}
\vspace*{-0.5cm}
\begin{center}
\large{
\vspace*{0.4cm}
\includegraphics[scale=0.20]{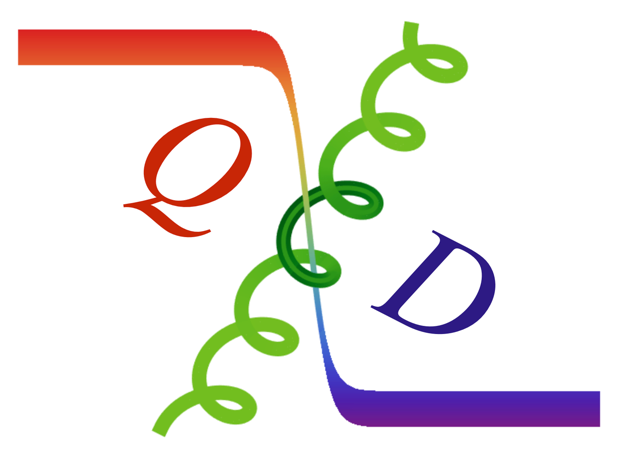}\\
\vspace*{0.4cm}
($\chi$QCD Collaboration)
}
\end{center}
}
\affiliation{
$^{1}$\mbox{Department of Physics and Astronomy, University of Kentucky, Lexington, KY 40506, USA}\\
$^{2}$\mbox{Maryland Center for Fundamental Physics, University of Maryland, College Park, Maryland 20742, USA}\\
$^{3}$\mbox{Nuclear Science Division, Lawrence Berkeley National Laboratory, Berkeley, CA, 94720, USA}\\
$^{4}$\mbox{Center for Theoretical Physics, Massachusetts Institute of Technology, Cambridge, MA 02139, USA}
}

\begin{abstract}
We present the 1-loop renormalization of the energy momentum tensor using the  overlap fermion and a HYP-smeared Iwasaki gauge action. 
We also calculate the 1-loop matching coefficient that convert the lattice simulation results renormalized in the RI/MOM scheme to the $\overline{\textrm{MS}}$ scheme.
 The dependence of the renormalization on the gauge action and the number of HYP smearing steps are also investigated.  \end{abstract}

\maketitle

\section{Introduction}

The energy momentum tensor (EMT) of QCD is central to our understanding of both the nucleon momentum as well as the nucleon angular momentum, particularly in understanding how the nucleon's momentum structure is built from the contributions of quark and gluon fields. Nearly two decades ago, Ji developed a gauge and frame independent decomposition of the proton spin where the quark and gluon augular momenta are constructed from the EMT of the Belinfante form~\cite{Ji:1996ek}: 

\bea
J^i_{Q,G}&=&\langle P,S | \int d^3x \, \epsilon^{ijk} x^j {\mathcal T}^{\{0k\}}_{Q,G} | P,S \rangle, \\
&&{\mathcal T}^{\{0k\}}_Q=\frac{1}{4}\bar{\psi}\gamma^{\{0}\overleftrightarrow{D}^{k\}}\psi,\ {\mathcal T}^{\{0k\}}_G=\epsilon^{klm}E^l B^m,\label{eq:operator}
\eea
where $\epsilon^{ijk}$ is the Levi-Civita tensor, ${\mathcal T}_{Q,G}$ is the quark/gluon EMT, and $E$ and $B$ are the color- electric and magnetic fields. The covariant derivative $D^\mu=\partial^\mu - igA^\mu$, and ``$\{\cdots\}$" in the superscript denotes the symmetrization of indices. Another related decomposition is that of the momentum fraction in the proton
\bea
\langle x \rangle_{Q,G}&=&\langle P,S | \int d^3x \, {\mathcal T}^{\{0i\}}_{Q,G} | P,S \rangle/P_i, 
\eea
where $p_i$ is the momentum of the proton along the spatial direction $i$.

Using Lorentz covariance, the matrix element of the EMT can be parametrized as
\bea
&&\langle P',S' |\int d^3x \, {\mathcal T}^{\{\mu\nu\}}_{Q,G} | P,S \rangle= \frac{1}{2} \bar{U}(P',S') 
\nonumber\\
  &&\qquad\big[T^{Q,G}_1(q^2)\gamma^{\{\mu}\bar{P}^{\nu\}}+ \frac{i}{2M}T^{Q,G}_2(q^2)\bar{P}^{\{\mu}\sigma^{\nu\}\alpha} q_{\alpha}\nonumber\\
  \nonumber\\
  &&\qquad\quad+ \frac{1}{M} T^{Q,G}_3(q^2) (q^{\mu} q^{\nu}-g^{\mu\nu}q^2)+M T^{Q,G}_4 g^{\mu\nu}\big] U(P,S) ,\nonumber\\
\eea
where $\bar{P}=(P+P')/2$, $q=P-P'$, $M$ is the nucleon mass and $T^{Q,G}_i(q^2)$ are frame independent form factors which can be related to both the momentum and angular momentum fractions  by~\cite{Ji:1996ek},
\bea
\langle x \rangle_{Q,G}=T^{Q,G}_1(0),\ 
J_{Q,G}=\frac{1}{2}[T^{Q,G}_1(0)+T^{Q,G}_2(0)].
\eea

Lattice QCD is the only practical method for model-independent predictions of the above quark and gluon fractions.  After lattice simulation, a non-trivial matching from bare quantities under the lattice regularization to the $\overline{\textrm{MS}}$ scheme under dimensional regularization is still required. The computation is complicated by the fact that the quark and gluon EMT operators mix with each other as well as other gauge variant operators.  These additional mixings make the 1-loop calculation, especially the gluonic sector, non-trivial. The calculations \cite{Corbo:1989ps,Corbo:1989yj,Capitani:1994qn} employing Wilson fermions and Wilson gluons were completed even before Ji's decomposition.  Since then, however, there has been limited progress and the interests of the lattice community have been concentrated on the renormalization of the quark sector.  This is partially due to the fact that the gluonic matrix elements in the nucleon are noisy and only recently are these matrix elements becoming available~\cite{Deka:2013zha, Horsley:2012pz,Alexandrou:2016ekb}. For the quark sector, there do exist calculations for more complicated actions \cite{Horsley:2005jk} at 1-loop level, and non-perturbative renormalization schemes have been developed for the quark bilinear operators \cite{Martinelli:1994ty} which have been used in many recent calculations .

The recent capability of the computer clusters allows lattice QCD simulations to obtain the gluonic matrix elements in the nucleon with an uncertainty at the 20\% level or better (see Refs.~\cite{Horsley:2012pz,Deka:2013zha,Yang:2016plb}). This new precision has increased the relevancy of computing the renormalization and mixing of the gluon operators. The full calculation has been revisited in Ref. \cite{Glatzmaier:2014sya} for Wilson fermions and a gluon operator defined by the overlap fermion action \cite{Alexandru:2008fu}. The focus of this work is both on the overlap fermion \cite{Neuberger:1997fp} as well as the clover definition~\cite{Capitani:1994qn} of the gluon operator with several HYP~\cite{Hasenfratz:2001tw}smearing steps.

This  paper is organized as follows, in section~\ref{sec:numberial}, we review the basic matching strategy which connects the bare quantities computed on the lattice to those renormalized under the $\overline{\textrm{MS}}$ scheme.  We also describe our approach to compute the 1-loop integrals numerically with a lattice regularization. A brief discussion of the Feynman rules for lattice perturbative theory is presented at the end of section~\ref{sec:numberial}. In the interest of brevity, the lengthy Feynman rules which have been listed in previous references are not reported here. Instead, we list the references and the corresponding equations where they can be found. In the first and second part of Section~\ref{sec:result}, we present the renormalization of the quark and gluon self-energies, and the 2x2 mixing matrix of the off-diagonal quark and gluon EMT. In section~\ref{sec:summary}, we list the final results and end the paper with a short discussion regarding the non-perturbative matching calculation.

\section{Numerical details}\label{sec:numberial}
\subsection{Matching from the bare lattice quantity to that in the $\overline{\textrm{MS}}$ scheme}

The strategy of matching the bare lattice quantity to that in the $\overline{\textrm{MS}}$ scheme described in Ref.~\cite{Capitani:2002mp} is to calculate the 1-loop matrix elements on the lattice as well as in the continuum,
\bea
\langle p|O_i^{lat}|p\rangle&=&\sum_j R^{lat}_{ij} \langle p|O_j^{tree}|p\rangle\nonumber\\
&=&\sum_j \Big(\delta_{ij}+\frac{g^2_0}{16\pi^2}\big (-\gamma_{ij} \textrm{log}(a^2p^2)  \nonumber\\
&&\quad +B^{lat}_{ij}(a^2p^2))\Big)\langle p|O_j^{tree}|p\rangle+O(g^4_0),\nonumber\\
 \langle p|O_i^{\overline{\textrm{MS}}}|p\rangle &=&\sum_j R^{\overline{\textrm{MS}}}_{ij} \langle p|O_j^{tree}|p\rangle\nonumber\\
&=&\sum_j \Big(\delta_{ij}+\frac{g^2_{\overline{\textrm{MS}}}}{16\pi^2}\big (-\gamma_{ij} \textrm{log}(\frac{p^2}{\mu^2}) +B^{\overline{\textrm{MS}}}_{ij})\Big) \nonumber\\
&&\quad \langle p|O_j^{tree}|p\rangle+O(g^4_{\overline{\textrm{MS}}}),
\eea
where $R_{ij}$ are the mixing coefficients at the quantum level,  $p^2\ll (\pi/a)^2$ is the infrared cutoff and the finite piece $B^{lat}_{ij}$ with the lattice regularization can be expanded as a polynomial in $a^2p^2$, $B^{lat}_{ij}(a^2p^2)=B^{(0)}_{ij}+O(a^2p^2)$. The exact value of $B^{(0)}_{ij}$ is sensitive to the lattice quark and gluon actions employed in the calculations.

Then we can convert the lattice matrix element into that in the $\overline{\textrm{MS}}$ scheme with
\bea
\langle p|O_i^{\overline{\textrm{MS}}}|p\rangle&=&\sum_{ij} R^{\overline{\textrm{MS}}}_{ik} (R^{lat}_{kj})^{-1} \langle p|O_j^{lat}|p\rangle \nonumber\\
&=&\sum_{ij} Z^{\overline{\textrm{MS}}}_{ij} \langle p|O_j^{lat}|p\rangle +O(a^2p^2),
\eea
where $Z^{\overline{\textrm{MS}}}_{ij}=\delta_{ij}+\frac{g^2_0}{16\pi^2}\big(\gamma_{ij} \textrm{log}(a^2\mu^2) +B^{\overline{\textrm{MS}}}_{ij}-B^{(0)}_{ij}\big)+O(g^4_0)$ is the effective renormalization matrix of the lattice matrix elements in the $\overline{\textrm{MS}}$ scheme. We ignored the difference between $g^2_0$ and $g^2_{\overline{\textrm{MS}}}$ since it is of the higher order in $g^2_0$, and will just use the notation $g^2$ in the following discussion.
The residual finite piece $B^{\overline{\textrm{MS}}}_{ij}-B^{(0)}_{ij}$ is generally non-zero, and introduces a $O(g^2_0)$ corrections on $\langle p|O_j^{lat}|p\rangle$ besides the standard $O(a^2)$ corrections.

The logic of the non-perturbative lattice renormalization is similar \cite{Martinelli:1994ty}. The matrix element $\langle p|O_i^{lat}|p\rangle$ can be calculated non-perturbatively, and decomposed into the parts proportional to the tree level matrix elements $\langle p|O_j^{tree}|p\rangle$, then the coefficients are the non-perturbative results of $R^{lat}_{ij}(a^2p^2)$. The calculation is repeated in the continuum perturbatively with higher loops corrections to obtain more accurate $R^{\overline{\textrm{MS}}}_{ij}$ as a function of the IR regulator $p^2$. Then one can combine $R^{lat}_{ij}(a^2p^2)$ and $R^{\overline{\textrm{MS}}}_{ij}$ to obtain $Z^{\overline{\textrm{MS}}}_{ij}$ as a function of $a^2p^2$, and an extrapolation of $a^2p^2$ is applied on $Z^{\overline{\textrm{MS}}}_{ij}(a^2p^2)$ to minimize the $O(a^2)$ corrections.

Such a strategy is equivalent to calculating the lattice renormalization constant $Z^{lat}_{ji}(\mu_R,a^{-1})$ under the RI/MOM scheme,
\bea
\sum_i Z^{lat}_{ji}(\mu_R,a^{-1})\langle p|O_i^{lat}|p\rangle|_{p^2=\mu^2_R}&=& \langle p|O_j^{tree}|p\rangle
\eea

and convert it into that under the $\overline{\textrm{MS}}$ scheme,
\bea
Z^{\overline{\textrm{MS}}}_{ij}=\sum_k R^{\overline{\textrm{MS}}}_{ik}Z^{lat}_{kj},
\eea
since $Z^{lat}_{ji}(\mu_R,a^{-1})$ is just $(R^{lat}_{kj}(a^2p^2)|_{p^2=\mu^2_R})^{-1}$.

\subsection{Calculation strategy}\label{sec:strategy}

For the 1-loop continuum perturbative theory (CPT) calculation, we use the newest version of the mathematica package, package-X \cite{Patel:2015tea}. This package can provide the analytic expressions of the Lorentz covariant integrations under dimensional regularization with very good performance. Compared with version 1.0, package-X version 2.0 (currently in beta) can handle integrations with repeated denominator factors (like $(p^2+m^2)^2$).  As a result, it is capable of handling the $\xi$-dependent part efficiently.

For lattice perturbative theory (LPT), an algorithm to obtain high precision 1-loop results has been developed \cite{Luscher:1995zz}.  We do not employ this algorithm here since such high precision is not necessary for the renormalization of the lattice simulation we study.  For example, if we consider a case where the lattice spacing is $a\sim0.1$ fm, then $g^2\sim3$ so that the finite piece in the 1-loop correction is proportional to is $\alpha_s/(4\pi)\sim0.02$. If we suppose the finite pieces contributing to the two-loop correction is as large as the 1-loop correction, we can expect it to contribute to the systematic uncertainty at the order of $4\times10^{-4}$. The precision of a modern lattice simulation for the matrix elements in the proton is, at best, at the 0.5\% level. So $B_{QQ}$  with around 0.01 uncertainty (or 0.02\% in the renormalization constant when the factor $\alpha/(4\pi)$ is included) is well within the precision requirements of modern lattice simulations. Because of this, many numerical integrators on current market will satisfy our recision requirements at the 1-loop level. We have found that a fast and desirable choice is the numerical integrator in Mathematica using the \textit{ClenshawCurtisOscillatoryRule} option. In practical tests, it performs integrations nearly 10 times faster than the well-known Monte Carlo integrator, Vegas.


Let us take the case of quark EMT operator
\bea
{\mathcal T}^{\{\mu\nu\}}_{Q,(0)}=\bar{\psi}(p)(\frac{\gamma^{\mu}p^{\nu}+\gamma^{\nu}p^{\mu}}{2}-\frac{1}{4}g^{\mu\nu}\dslash{p})\psi(-p)
\eea
 with the Wilson fermion and Wilson gauge action as an example. It is trivial to confirm from CPT that the bare quark operator ${\mathcal T}^{\{\mu\nu\}}_{q,bare}$ under the lattice regularization has the following form,
\bea\label{eq:quark_example}
{\mathcal T}^{\{\mu\nu\}}_{Q,bare}&=&\Big(1+\frac{g^2C_F}{16\pi^2}[A\,\textrm{log}(a^2p^2)+B]\Big){\mathcal T}^{\{\mu\nu\}}_{Q,(0)}\nonumber\\
&&-\frac{g^2C_F}{16\pi^2}C \frac{p^{\mu}p^{\nu}}{p^2}\bar{\psi}\dslash{p}\psi +O(g^4).
\eea
where $A=8/3$ and $C=4/3-\xi$ are universal in both CPT and LPT, and Ref. \cite{Capitani:2002mp} provides that $B=-3.16486+\xi$ for the Wilson fermion and gluon action. For the case of $\mu\neq\nu$, one can contract Eq.~(\ref{eq:quark_example}) with $\gamma_\nu$ to obtain the renormalization factor as:
\bea\label{eq:value_ref}
Z(p)&\equiv&\frac{\textrm{Tr}[\gamma_\nu.{\mathcal T}^{\{\mu\nu\}}_{Q,(0)}]}{\textrm{Tr}[\gamma_\nu.{\mathcal T}^{\{\mu\nu\}}_{Q,bare}]}=1-\frac{g^2C_F}{16\pi^2}I_\nu(p),\nonumber\\
I_\nu(p)&=& \frac{8}{3}\textrm{log}(a^2p^2)-3.16486+\xi-\frac{4-3\xi}{3}\frac{2(p_\nu)^2}{p^2},\nonumber\\
\eea
where we have amputated the external legs of the EMT, and 
the index $\nu$ is not summed.

In this work, we do the numerical integration $I_1(p)$ for 16 external momenta, $p_{(i)}=(0.003*1.1^{i},0,0,0)$ ($i=0,...,15$), and fit the results to the following functional form (an overall factor $g^2C_F$ has been dropped),
\bea
f_0(a^2p^2)=\frac{1}{16\pi^2}[A_0\textrm{Log}(a^2p^2)+B_0],
\eea 
an analytic computation of $A_0$ gives a value for $A_0$ to be 8/3. The value we obtained is 2.6666.

After computing all integrations, we fit the results to the following functional forms,
\bea
f_1(a^2p^2)&=&\frac{1}{16\pi^2}[\frac{8}{3}\textrm{Log}(a^2p^2)+B_1+C_1 a^2p^2],\nonumber\\
f_2(a^2p^2)&=&\frac{1}{16\pi^2}[\frac{8}{3}\textrm{Log}(a^2p^2)+B_2+C_2 a^2p^2+D_2 a^4p^4],\nonumber\\
f_3(a^2p^2)&=&\frac{1}{16\pi^2}[\frac{8}{3}\textrm{Log}(a^2p^2)+B_3+C_3 a^2p^2+D_3 a^4p^4\nonumber\\
&&\quad+E_3 a^6p^6],
\eea
and estimate the finite piece by taking $B_2$ as the central value. The uncertainty of each finite peice is estimated in two ways.  First, by the variance of $B_{1,2,3}$ and the second by the averaged bias of the fit,
\bea
\sigma=\sqrt{\frac{\sum_{i=1,2,3;j=0,...,N-1} (f_i(p_{(j)})-I(p_{(j)}))^2}{3N}}
\eea
where $N$=16 is the number of independent momenta values used. We added each uncertainty in quadrature, and took this value to be the final uncertainty estimate.   The final value we obtain is $B_0=-3.1654(12)$, which agrees perfectly with the value quoted in Eq.~(\ref{eq:value_ref}). In addition, we computed this value using the momenta $p_{(i)}=(0.003*1.1^{i},0,0.003*1.1^{i},0)$ ($i=0,...,15$), and obtained a consistent estimate -3.1646(6). 

As an additional check, we consider the mixing coefficient $C=4/3-\xi$. We can repeat the calculation with the momenta $p_{(i)}=(0.003*1.1^{i},0.003*1.1^{i},0,0)$, ($i=0,...,15$) and two values of the gauge parameter $\xi$= 0, 1. Since only one gluon propagator is involved in the loop integrations, we expect that the dependence on $\xi$ be linear. The estimate of the finite piece is -4.4986(6)+2.0004(8)$\xi$ and then that of $C$ is $1.3336(8)-1.0002(4)\xi$, again consistent with the accurate value $4/3-\xi$.

In all computations that follow, we keep two digits after the decimal point.  This is sufficient given the present precision of the Lattice QCD simulation.

\subsection{Feynman diagrams and rules}\label{sec:feynman_rules}.

\begin{figure}[!h]
  \includegraphics[scale=0.25]{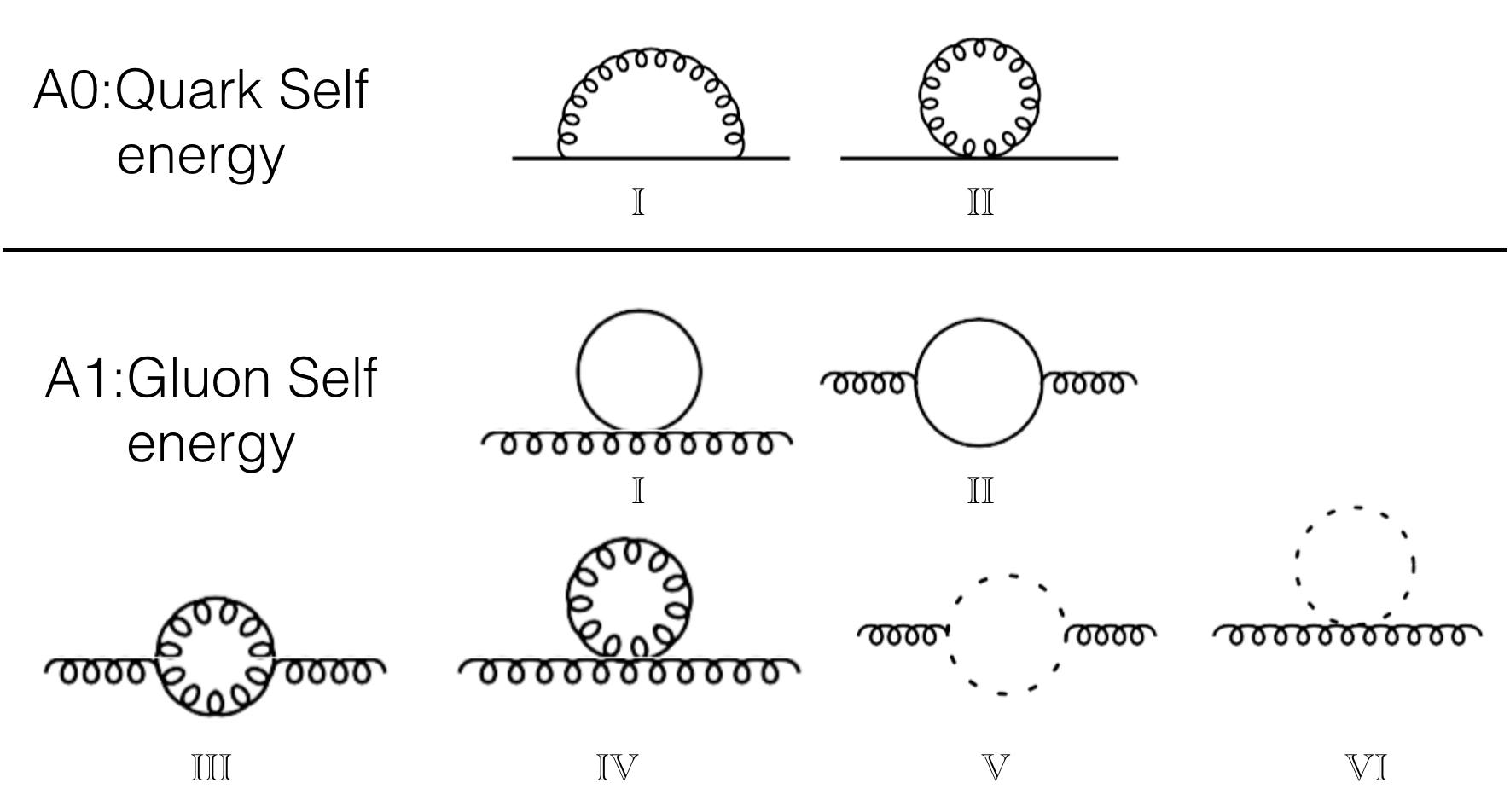}
 \caption{The diagrams of the quark/gluon self-energies. Only the diagrams A0.I, A1.II, A1.III, A1.V are left in the continuum calculation.}\label{fig:self}
\end{figure}

\begin{figure*}[ht]
  \includegraphics[scale=0.4]{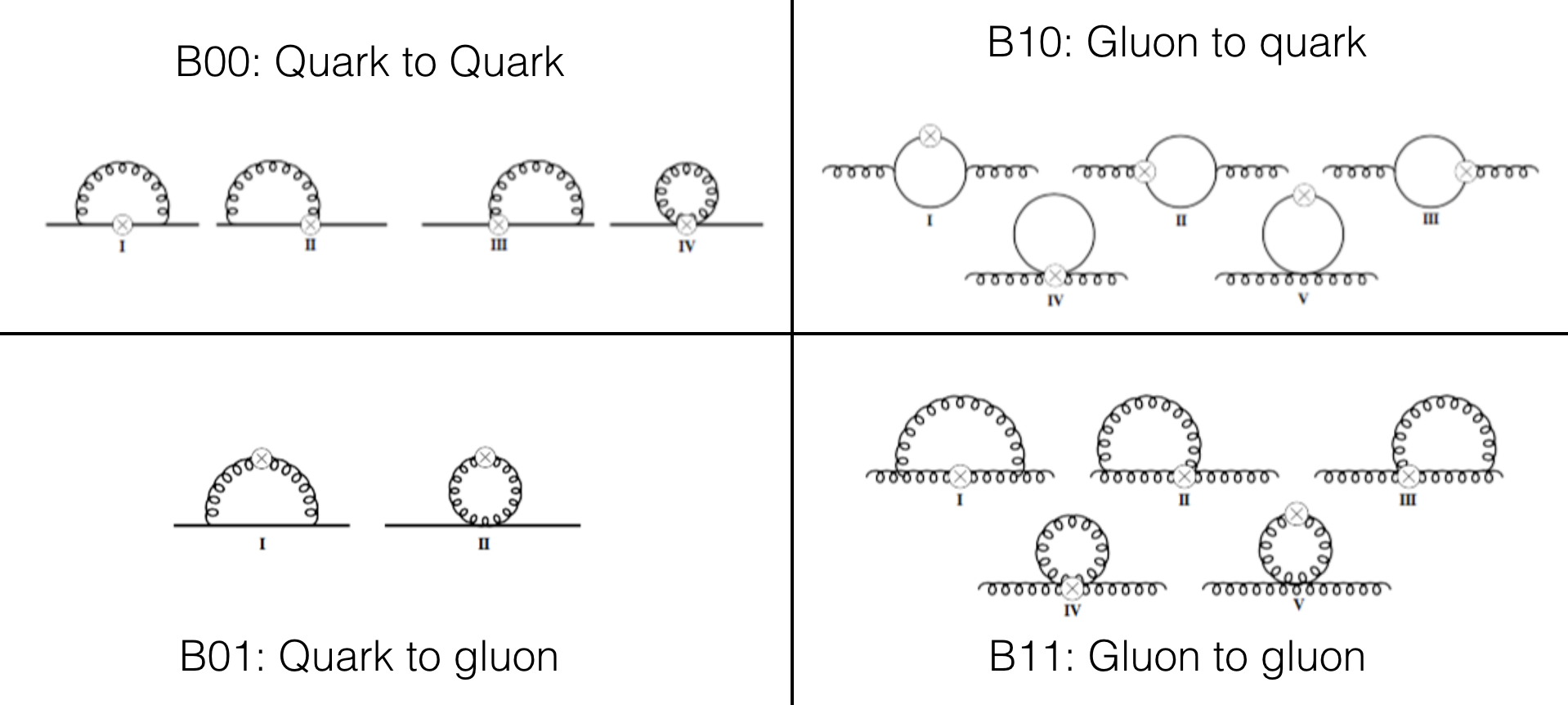} 
 \caption{The diagrams of the one loop corrections of the EM tensor. Four sub-panels for the quark operator renormalization, the mixing from the gluon operator to that of quark, and so on. Note that in the ConPT, the diagrams B00.IV, B10.IV, B10.V, B01.II are not exist and the contribution from B11.IV and B11.V are zero.}\label{fig:mixing}
\end{figure*}

The Feynman diagrams needed for the 1-loop calculation of the self-energy and the EMT vertices corrections are listed in Fig.~\ref{fig:self} and ~\ref{fig:mixing} respectively. Note that in the continuum, the diagrams A0.II, A1.I, A1.VI, B00.IV, B01.IV, B01.V, and B10.II do not exist and the contributions from the diagrams A1.IV, B11.IV and B11.V are zero (since they are independent of the momentum on the external legs).

The LPT Feynman rules used in our computations can be found in the following references:

1. \textit{Wilson fermion}: \qquad Eq. (5.74), (5.76) and (5.78) in Ref.~\cite{Capitani:2002mp}.

2. \textit{Overlap fermion}:\qquad Eq. (95-100) in Ref.~\cite{Horsley:2004mx}

3. \textit{Chiral fermion $D_c$}: \qquad The Feynman rules of the propagator, $qqg$ and $qqgg$ vertices with the chiral fermion 
$D_c\equiv D_{ov}/(1-1/(2\rho)D_{ov})$
which satisfies $\{\gamma_5,D_c\}=0$ are:
\begin{widetext}
\bea
S^{D_c}&=&S^{D_{ov}}-\frac{1}{2\rho}\ , \nonumber\\
V^{D_c}_{1\mu}(p_1,p_2)&=&\frac{1}{1-\frac{1}{2\rho}D_{ov}(p_1)}V^{D_{ov}}_{1\mu}(p_1,p_2)\frac{1}{1-\frac{1}{2\rho}D_{ov}(p_2)}\ , \nonumber\\
V^{D_c}_{2\mu\nu}(p_1,p_2,k_1,k_2)&=&\frac{1}{1-\frac{1}{2\rho}D_{ov}(p_1)}\left[V^{D_{ov}}_{2\mu}(p_1,p_2,k_1,k_2)\right.\nonumber\\
&& \quad -\frac{1}{4\rho}V^{D_{ov}}_{1\mu}(p_1,p_1+k_1)\frac{1}{1-\frac{1}{2\rho}D_{ov}(p_1+k_1)}V^{D_{ov}}_{1\nu}(p_1+k_1,p_2)\nonumber\\
&& \quad \left.-\frac{1}{4\rho}V^{D_{ov}}_{1\nu}(p_1,p_1+k_2)\frac{1}{1-\frac{1}{2\rho}D_{ov}(p_1+k_2)}V^{D_{ov}}_{1\mu}(p_1+k_2,p_2)\right]\frac{1}{1-\frac{1}{2\rho}D_{ov}(p_2)}\ .
\eea
\end{widetext}
where $S^{D_{ov}}=1/D_{ov}$, $V^{D_{ov}}_{1\mu}(p_1,p_2)$ and $V^{D_{ov}}_{2\mu}(p_1,p_2,k_1,k_2)$ are the Feynman rules for the propagator, $qqg$ and $qqgg$ vertices respectively with the standard overlap fermion $D_{ov}$ ~\cite{Horsley:2004mx}. 

One can check that the quark-glue mixings with the $D_c$ action are identical to that with the $D_{ov}$ action, and the renormalization of the $D_c$ quark bilinear operators are the same as the overlap in the $ap\ll1$ limit. However we have found that the finite pieces from diagrams involving a quark loop contributing to the gluon self-energy are different for the $D_c$ and $D_{ov}$ cases. 

4. \textit{Wilson and Iwasaki gluon}:\qquad The Feynman rules of the generic $O(a^2)$ improved gluon are listed in Eq. (A.8-A.31) of Ref.~\cite{Weisz:1983bn}. The Wilson gluon action without any $O(a^2)$ improvement corresponds to the case $c_0=1,c_1=0$ and the Iwasaki action which has such an improvement corresponds to $c_0=1-8c_1,c_1=-0.331$. We confirmed that the 4-gluon vertex in Eq. (A.19) of Ref.~\cite{Weisz:1983bn} is equivalent to that in Eq. (5.28) in Ref.~\cite{Capitani:2002mp}, with proper combination in Eq. (A.7) of Ref.~\cite{Weisz:1983bn}.

5. \textit{Covariant ghost}:\qquad Eq. (5.66-5.68) in Ref.~\cite{Capitani:2002mp}.

6. \textit{HYP smearing}:\qquad Equations in Sec. 3 of \cite{Hasenfratz:2001tw}. 
For the lattice simulation, the gluon field was fattened with one step of the HYP smearing for both the quark EMT operator and the inversion of the quark propagators. Five steps of HYP smearing were used for the glue EMT operator. We employed the Iwasaki gauge action for all calculations, and we note here that no reweighting was applied to the configurations.  For the perturbative matching calculations, since the gauge action was left unchanged, we apply the projection operator derived in Ref. \cite{Hasenfratz:2001tw}.  One step of HYP smearing corresponds to applying this projection operator once at each quark-gluon vertex, and five times for the glue EMT operator. No HYP smearing was applied to the Feynman rules for the 3-gluon and 4-gluon vertices of the gluon action, due to the fact that the gluon action is not HYP smeared.

7. \textit{Quark and gluon EMT operator}:\qquad  The expressions for both the tree level and $O(g)$ vertices of the quark and gluon operators can be found in Eq. (A.5-A.8) of Ref.~\cite{Capitani:1994qn}. We note that the $O(a)$ corrections outlined in Eq. (A.5-A.8) of Ref.~\cite{Capitani:1994qn} were dropped in this work since we do not apply any improvement to these operators. The Feynman rule for the quark $O(g^2)$ vertex is,
\bea\label{eq:q2}
&&O^q_{\mu\nu}|_{g^2}(p,p,k,k)=\nonumber\\
&&\quad-ia/2g^2 \int^{\pi}_{-\pi} \bar{\psi}(p)\gamma_{\mu} \textrm{sin}(ap_{\nu})A_{\nu}(k)A_{\nu}(-k)\psi(p).
\eea

The corresponding Feynman rules for the twist-two EMT gluon operator are more complex, however.  The Feynman rules of the gluon EMT operator can be expressed as,
\bea
O^G_{\mu\nu}&=&G_{\mu\rho}G_{\rho\nu}=O^{G,(0)}_{\mu\nu}+g^1 O^{G,(1)}_{\mu\nu} +g^2 O^{G,(2)}_{\mu\nu} +O(g^3)\nonumber\\
\eea
with
\bea
O^{G,(0)}_{\mu\nu}&=&G^{(0)}_{\mu\rho}G^{(0)}_{\rho\nu},\\
O^{G,(1)}_{\mu\nu}&=&G^{(0)}_{\mu\rho}G^{(1)}_{\rho\nu}+G^{(1)}_{\mu\rho}G^{(0)}_{\rho\nu},\\
O^{G,(2)}_{\mu\nu}&=&G^{(0)}_{\mu\rho}G^{(2)}_{\rho\nu}+G^{(1)}_{\mu\rho}G^{(1)}_{\rho\nu}+G^{(2)}_{\mu\rho}G^{(0)}_{\rho\nu},\label{eq:g4}
\eea
where $G_{\mu\nu}=\sum_i g^{i}G^{(i)}_{\mu\nu}$. The expression of $O^{G,{(0)}}_{\mu\nu}$ and $O^{G,{(1)}}_{\mu\nu}$ are listed in Ref.~\cite{Capitani:1994qn} and then we can deduce the form of  $G^{(0,1)}_{\mu\nu}$ as
\begin{widetext}
\bea\label{eq:g1}
G^{(0),a}_{\mu\nu}(q_1)&=&\frac{1}{a}(A^a_{\mu}(q_1)\textrm{cos}\frac{q_{1\mu}}{2}\textrm{sin}q_{1\nu}-A^a_{\nu}(q_1)\textrm{cos}\frac{q_{1\nu}}{2}\textrm{sin}q_{1\mu})\ ,\\
G^{(1),a}_{\mu\nu}(q_1,q_2)&=&f_{abc}A^b_{\mu}(q_1)A^c_{\nu}(q_2)\big[ \frac{1}{2}(\textrm{cos}\frac{q_{1\mu}}{2}-\textrm{cos}\frac{q_{1\mu}+2q_{2\mu}}{2})(\textrm{cos}\frac{q_{2\nu}}{2}-\textrm{cos}\frac{q_{2\nu}+2q_{1\nu}}{2}) -\textrm{cos}\frac{q_{1\mu}+2q_{2\mu}}{2}\textrm{cos}\frac{q_{2\nu}+2q_{1\nu}}{2}\big]\nonumber\\
&&\quad +\frac{1}{2}f_{abc}\big(-A^b_{\mu}(q_1)A^c_{\mu}(q_2)\textrm{sin}\frac{q_{1\mu}+2q_{2\mu}}{2}\textrm{sin}\frac{q_{2\nu}}{2}+A^b_{\nu}(q_1)A^c_{\nu}(q_2)\textrm{sin}\frac{q_{1\nu}+2q_{2\nu}}{2}\textrm{sin}\frac{q_{2\mu}}{2}\big)\ .
\eea
\end{widetext}
where the superscipt $a$ in $G^{(0,1),a}$ is the color index. But $G^{(2)}_{\mu\nu}$ is very complicated and is not shown in Ref.~\cite{Capitani:1994qn}. 

The gluon $O(g^2)$ vertex we need is $O^{G,(2)}_{\mu\nu}=G^{(0)}_{\mu\rho}G^{(2)}_{\rho\nu}+G^{(1)}_{\mu\rho}G^{(1)}_{\rho\nu}+G^{(2)}_{\mu\rho}G^{(0)}_{\rho\nu}$. The vertex has four gluon external legs, appearing in the tadpole diagram B11.IV of Fig.~\ref{fig:mixing} which contributes to the vertex correction of the gluon EMT operator. We can't obtain its whole contribution since we don't know the expression of $G^{(2)}_{\mu\nu}$. However, the general color structure of the terms in the $O^{G,2}_{\mu\nu}$ expression are known \cite{Capitani:1994qn,Capitani:2002mp},
\begin{widetext}
\bea
G^{(1)}_{\mu\rho}(q_1,q_2)G^{(1)}_{\rho\nu}(q_3,q_4)&=& f_{abc}f_{ade} F_{1,\alpha\beta\gamma\delta\mu\nu}(q_1,q_2,q_3,q_4) A_{\alpha}^b(q_1)A_{\beta}^c(q_2)A_{\gamma}^d(q_3)A_{\delta}^e(q_4),\nonumber\\
G^{(0)}_{\mu\rho}(q_1)G^{(2)}_{\rho\nu}(q_2,q_3,q_4)&=&\big(f_{abc}f_{ade} F_{2,\alpha\beta\gamma\delta\mu\nu}(q_1,q_2,q_3,q_4)\nonumber\\
&&+\textrm{Tr}[\{T^b,T^c\}\{T^d,T^e\}+(c\leftrightarrow e)+(d\leftrightarrow e)] 
   F_{3,\alpha\beta\gamma\delta\mu\nu}(q_1,q_2,q_3,q_4)\big)\nonumber\\
&&A_{\alpha}^b(q_1)A_{\beta}^c(q_2)A_{\gamma}^d(q_3)A_{\delta}^e(q_4).   
\eea
\end{widetext}
where $F_1$ can be obtained from Eq.~(\ref{eq:g1}) and we need to estimate the contribution from $F_2$ and $F_3$. The contribution of the $F_1$ and $F_2$ are proportional to $N_c$ and that of $F_3$ is proportional to $\frac{2N_c^2-3}{N_c}$.

A. The contribution of the $F_2$ term: In the Wilson gluon case, the joint contribution of $F_1$ and $F_2$ is known~\cite{Capitani:1994qn}, and the contribution of the $F_1$ term can be calculated. We can take the difference to get the contribution of the $F_2$ term and it is 0.389 of that of the $F_1$ term. 
Supposing the $O(a^2)$ improvement of the gluon action and the HYP smearing does not change this ratio, we can obtain an estimate of the contribution of the $F_2$ term by multiplying the same factor 0.389 on the $F_1$ term contributions. 

B. The contribution of the $F_3$ term: The four gluon vertex in the gluon action also has a part with the same color structure as that in front of $F_3$, and it contributes to the gluon self-energy (the corresponding Feynman diagram is the diagram A1.IV of Fig.~\ref{fig:self}). That contribution to the gluon self-energy, is exactly -1/2 of the $F_3$ term contribution in the gluon vertex correction. If we use the same assumption as the $F_2$ case, we can calculate those contribution in the glue self-energy case but with the improved action (and/or applying the HYP smearing on the external legs of the $O(g^4)$ term of the action), and multiply -2 to estimate the contribution of the $F_3$ term in the gluon vertex correction.

The uncertainties in the $F_2$ and $F_3$ terms are taken to be 100\% and are added together in quadrature.

\section{Results}\label{sec:result}

\subsection{Self energies}

In CPT, the ratio 
\bea
R_Q^{\overline{\textrm{MS}}}\equiv \frac{\langle \bar{\psi}^{\overline{\textrm{MS}}} \psi^{\overline{\textrm{MS}}}\rangle}{\langle\bar{\psi}^{tree} \psi^{tree}\rangle},\ 
R_G^{\overline{\textrm{MS}}}\equiv \frac{\langle A^{\overline{\textrm{MS}}} A^{\overline{\textrm{MS}}}\rangle}{\langle A^{tree} A^{tree}\rangle},
\eea
 for the quark and gluon self-energy are
\bea
R^{\overline{\textrm{MS}}}_{Q}&=&1+\frac{g^2C_F}{16\pi^2}[(1-\xi)\textrm{log}(\mu^2/p^2)+1-\xi]+O(g^4),\nonumber\\
R^{\overline{\textrm{MS}}}_{G}&=&1+\frac{g^2}{16\pi^2}[N_f\big(\frac{2}{3}\textrm{log}(\mu^2/p^2)+\frac{10}{9}\big)\nonumber\\
&&\quad-N_c\big(R^{\overline{\textrm{MS}}}_{G,g}+R^{\overline{\textrm{MS}}}_{G,c}\big)],
\eea
where 
$R^{\overline{\textrm{MS}}}_{G,g/c}$ is the contribution from the gluon/ghost diagram respectively. For the part proportional to $p^{\mu}p^{\nu}$ in the 1-loop matrix element,
\bea
R^{\overline{\textrm{MS}}}_{G,g}&=&-\frac{g^2}{16\pi^2}[\frac{1}{6}\textrm{log}(\mu^2/p^2)+\frac{5}{18}],\nonumber\\
R^{\overline{\textrm{MS}}}_{G,c}&=&\frac{g^2}{16\pi^2}[\frac{11+3\xi}{6}\textrm{log}(\mu^2/p^2)+\frac{134-36\xi+9\xi^2}{36}].
\eea
and for the part proportional to $g^{\mu\nu}p^2$,
\bea
R^{\overline{\textrm{MS}}}_{G,g}&=&\frac{g^2}{16\pi^2}[\frac{1}{12}\textrm{log}(\mu^2/p^2)+\frac{2}{9}],\nonumber\\
R^{\overline{\textrm{MS}}}_{G,c}&=&\frac{g^2}{16\pi^2}[\frac{19+6\xi}{12}\textrm{log}(\mu^2/p^2)+\frac{116-36\xi+9\xi^2}{36}].
\eea
Combining all results above leads to a single term proportional to $g^{\mu\nu}p^2-p^{\mu}p^{\nu}$, as required by gauge sysmmetry,
\bea
R^{\overline{\textrm{MS}}}_{G,g}+R^{\overline{\textrm{MS}}}_{G,c}=\frac{10+3\xi}{6}\textrm{log}(\mu^2/p^2)+\frac{124-36\xi+9\xi^2}{36}.
\eea

The ratio $R^{lat}$ defined similarly can be expressed in terms of the loop integrations,
\bea
R^{lat}_Q&=&1-g^2C_F f_Q +O(g^4) \nonumber\\
R^{lat}_{G}&=&1-g^2N_f f^f_{G}- g^2N_c f_{G}+O(g^4),
\eea
where the loop integration $f_Q$ is defined by the projection operation,
\bea
f_Q&=&\frac{1}{4p_0}\textrm{Tr}[(I_{A0.I}(p)+I_{A0.II}(p)).\gamma_0].\label{eq:self_q}
\eea
Here $I_{X}$ is the loop integration based on the subpanel $X$ in Fig.~\ref{fig:self} and $p=(\epsilon,0,0,0)$ with several different values of $\epsilon$ to apply the scheme described in Sec.~\ref{sec:strategy}.
However, the $ f^f_{G}$ and $ f_{G}$ case are less straightforward. In these cases we are interested in the pieces proportional to $g^{\mu\nu}$ but it is known that a power divergence appears here which affects the precision of the finite piece we want. Suppose the following function $f_(p^2)$ can be written as
\bea
f(p^2)=\frac{\Delta}{p^2}+A \textrm{Log}(p^2)+B
\eea
where $\Delta$, $A$ and $B$ are constants, then we can fit $\frac{\partial}{\partial p^2}(p^2 f(p^2))$ instead of $f(p^2)$ itself with the strategy described in the previous section, to extract $A$ and $B$ without touching $\Delta$ which includes the power divergence.
 With these manipulations, we find
\bea
f^f_G&=&\frac{\partial}{\partial p^2}\big(I^{11}_{A1.I}(p)+I^{11}_{A1.II}(p)\big)-\frac{1}{16\pi^2}\frac{2}{3},\nonumber\\
f_G&=&\frac{\partial}{\partial p^2}\sum_{i=III,IV,V,VI}I^{11}_{A1.i}(p)+\frac{1}{16\pi^2}\frac{10+3\xi}{6},\label{eq:self_g}
\eea
where $I^{\rho,\tau}_{X}$ is the loop integration based on the subpanel $X$ of Fig.~\ref{fig:self} and the Lorentz indices of the external legs are $\rho$ and $\tau$.  The momentum components $p^{\rho}$ and $p^{\tau}$ on the external legs should be zero to avoid the mixing with the term proportional to $p^{\rho}p^{\tau}$.

The final renormalization constants in the $\overline{\textrm{MS}}$ scheme are,
\bea
Z^{\overline{\textrm{MS}}}_Q&=&1+\frac{g^2C_F}{16\pi^2}[(1-\xi)\textrm{log}(a^2\mu^2)+B_Q+1+3.79\xi]\nonumber\\
&&+O(a^2p^2)+O(g^4),\nonumber\\
Z^{\overline{\textrm{MS}}}_{G}&=&1+\frac{g^2}{16\pi^2}[N_f\big(\frac{2}{3}\textrm{log}(a^2\mu^2)+\frac{10}{9}+B^f_{G}\big)\nonumber\\
&&-N_c\big(\frac{10+3\xi}{6}\textrm{log}(a^2\mu^2) +B_G+\frac{31}{9}+6.60\xi)\big)]\nonumber\\
&&+O(a^2p^2)+O(g^4),
\eea
where $B_Q$, $B^f_G$ and $B_G$ are sensitive to the quark and gluon actions used. The values in different cases are listed in Table~\ref{tab:self}. In the Table~\ref{tab:self}, $B_G$ is split into two pieces 
\bea
B_G=B^t_G+\frac{2N_c^2-3}{24N^2_c}B^{s}_{G}
\eea
where the first term is the contribution from the color structure $2\delta^{cd} \textrm{Tr}[T^a,T^c][T^d,T^b]=\delta^{ab}N_c$ in the continuum, and the second term is that from the additional color structure $\frac{\delta^{cd}}{4!}\textrm{Tr}[\{T^a,T^b\}\{T^c,T^d\}+(b\leftrightarrow d)+(b\leftrightarrow d)]=\delta^{ab}\frac{2N_c^2-3}{24N_c}$ from the lattice regularization.

\begin{table}[htbp]
\begin{center}
\caption{\label{tab:self} The finite pieces $B_Q$, $B^f_G$ and $B_G$ in the self-energies in different cases. The rows are for different quark actions and the columns for the gauge actions. $B_G=B^t_G+\frac{2N_c^2-3}{24N^2_c}B^{s}_{G}$ are split to two parts with different color structures.}
\begin{tabular}{c|ccc|c}
\hline
&\multicolumn{3}{c|}{$B_Q$} & $B^f_G$\\
& Wilson & Iwasaki & Iwasaki$^{HYP}$  &\\
\hline
Wilson & \ 11.85  &\ \ 3.32 & -4.22  & -2.17\\
overlap & -21.50 & -13.58 & -7.56  &-0.72\\
$D_c$ & & & &\ 0.16\\
\hline
&\multicolumn{3}{c|}{$B_G$} &\\
   & Wilson & Iwasaki & -- \\
\hline
$B^t_{G}$  &\  -8.53  &  -0.05  &--  \\
$\frac{2N_c^2-3}{24N^2_c}$$B^s_{G}$ &  -10.97  & \ 6.61  &  --  \\
$B_{G}$ & -19.49 &  \  6.56 & --
\end{tabular}
\end{center}
\end{table}

\subsection{Mixings}

The tree-level form of the quark traceless EMT is simple while that of the gluon EMT is quite complicated,
\bea
\overline{{\mathcal T}}^{\{\mu\nu\}}_{Q,(0)}&=&\bar{\psi}(p)(\frac{\gamma^{\mu}p^{\nu}+\gamma^{\nu}p^{\mu}}{2}-\frac{1}{4}g^{\mu\nu}\dslash{p})\psi(-p),\label{eq:q_tree}\\
\overline{{\mathcal T}}^{\{\mu\nu\}}_{G,(0)}&=&A^a_{\rho}(p)A^a_{\tau}(-p)\nonumber\\
&&\big(-2p^{\mu}p^{\nu}g^{\rho\tau} + p^{\mu}p^{\rho}g^{\nu\tau}-p^2g^{\rho\mu}g^{\nu\tau}+p^{\tau}p^{\nu}g^{\rho\mu}\nonumber\\
&&\qquad+p^{\nu}p^{\rho}g^{\mu\tau}-p^2g^{\rho\nu}g^{\mu\tau}+p^{\tau}p^{\mu}g^{\rho\nu}\nonumber\\\label{eq:g_tree}
&&\qquad-g^{\mu\nu}(p^{\tau}p^{\rho}-p^2g^{\tau\rho})\ \big),
\eea
where $\mu$ and $\nu$ denote the external Lorentz indices of EMT. For the gluon operator $\overline{{\mathcal T}}^{\{\mu\nu\}}_{G,(0)}$, we focus on the coefficient of the term which does not vanish under the following physical conditions
\bea
p_{\rho}=p_{\tau}=0, p^2=0,
\eea
where $\rho$ and $\tau$ are the indices of the external legs~\cite{Collins:1994ee}. With these conditions, the only non-vanishing Lorentz structure is $-2p^{\mu}p^{\nu}g^{\rho\tau}$.

To avoid mixing with terms from the QCD equation of motion, the ratio $R^{lat/\overline{\textrm{MS}}}_{ij}$ for the off-diagonal pieces of the EMT are defined by the following equations ($\mu\neq\nu$),
\bea\label{eq:condition_off}
&\langle Q |\overline{{\mathcal T}}^{\{\mu\nu\}}_{Q}|Q\rangle|_{p_{\nu}=0}&=R_{QQ}\frac{\gamma_{\nu}p_{\mu}}{2},\nonumber\\
&\langle G,\rho |\overline{{\mathcal T}}^{\{\mu\nu\}}_{Q}|G,\tau\rangle|_{\rho=\tau\neq\mu,\nu,p_\rho=0}&=R_{GQ}(-2p_\mu p_\nu),\nonumber\\
&\langle Q |\overline{{\mathcal T}}^{\{\mu\nu\}}_{G}|Q\rangle|_{p_{\nu}=0}&=R_{QG}\frac{\gamma_{\nu}p_{\mu}}{2},\nonumber\\
&\langle G,\rho |\overline{{\mathcal T}}^{\{\mu\nu\}}_{G}|G,\tau\rangle|_{\rho=\tau\neq\mu,\nu,p_\rho=0}&=R_{GG}(-2p_\mu p_\nu),
\eea
where $|Q\rangle$ and $|G, \sigma\rangle$ are the quark and gluon states with the Lorentz index $\sigma$ respectively.  
They are equivalent to the renormalization conditions under the RI-MOM scheme which can be chosen to be~\cite{Capitani:1994qn},
\bea
&\langle Q |\overline{{\mathcal T}}^{\{\mu\nu\},R}_{Q}|Q\rangle|_{p_{\nu}=0,p^2=\mu^2_R}&=\frac{\gamma_{\nu}p_{\mu}}{2},\nonumber\\
&\langle G,\rho |\overline{{\mathcal T}}^{\{\mu\nu\},R}_{Q}|G,\tau\rangle|_{p^2=\mu^2_R}&=0,\nonumber\\
&\langle Q |\overline{{\mathcal T}}^{\{\mu\nu\},R}_{G}|Q\rangle|_{p^2=\mu^2_R}&=0,\nonumber\\
&\langle G,\rho |\overline{{\mathcal T}}^{\{\mu\nu\},R}_{G}|G,\tau\rangle|_{\rho=\tau\neq\mu,\nu,p_\rho=0,p^2=\mu^2_R}&=-2p_\mu p_\nu.
\eea
The $R^{T,lat/\overline{\textrm{MS}}}_{ij}$ of the trace-less diagonal pieces of the EMT can be defined similarly,
\bea\label{eq:condition_tr}
&\langle Q |\overline{{\mathcal T}}^{\{44\}}_{Q}|Q\rangle|_{p=(k,k,k,k)}&=R^T_{QQ}(\gamma_{4}p_{4}-\dslash{p}/4),\nonumber\\
&\langle G,3 |\overline{{\mathcal T}}^{\{44\}}_{Q}|G,3\rangle|_{p=(0,0,0,k)}\nonumber\\
&\quad\quad-\langle G,3 |\overline{{\mathcal T}}^{\{44\}}_{Q}|G,3\rangle|_{p=(k,0,0,0)}&=R^T_{GQ}(-2k^2),\nonumber\\
&\langle Q |\overline{{\mathcal T}}^{\{44\}}_{G}|Q\rangle|_{p=(k,k,k,k)}&=R^T_{QG}(\gamma_{4}p_{4}-\dslash{p}/4),\nonumber\\
&\langle G,3 |\overline{{\mathcal T}}^{\{44\}}_{G}|G,3\rangle|_{p=(0,0,0,k)}&\nonumber\\
&\quad\quad-\langle G,3 |\overline{{\mathcal T}}^{\{44\}}_{G}|G,3\rangle|_{p=(k,0,0,0)}&=R^T_{GG}(-2k^2),
\eea
where the superscript $T$ is added in $R^T_{ij}$, to distinguish it from the ratio $R_{ij}$ in the off-diagonal case. Note that we used a special condition for $R^T_{GG}$ to remove the unwanted parts proportional to $p^2$ and get the correct component, by taking the difference of two matrix elements with different external momenta.

The ratios $R^{\overline{\textrm{MS}}}$ and $R^{T,\overline{\textrm{MS}}}$ are the same due to rotation sysmmetry,
\begin{widetext}
\bea
R^{\overline{\textrm{MS}}}\equiv\left(\begin{array}{cc}
R^{\overline{\textrm{MS}}}_{QQ}&
R^{\overline{\textrm{MS}}}_{GQ}\\
R^{\overline{\textrm{MS}}}_{QG}&
R^{\overline{\textrm{MS}}}_{GG}
\end{array}\right)&=&
\left(\begin{array}{ll}
1-\frac{g^2C_F}{16\pi^2}[\frac{8}{3}\textrm{log}(\mu^2/p^2)+\frac{40-9\xi}{9}]&
0+\frac{g^2N_f}{16\pi^2}[\frac{2}{3}\textrm{log}(\mu^2/p^2)+\frac{4}{9}] \\
0+\frac{g^2C_F}{16\pi^2}[\frac{8}{3}\textrm{log}(\mu^2/p^2)+\frac{22}{9}] &
1-\frac{g^2N_f}{16\pi^2}[\frac{2}{3}\textrm{log}(\mu^2/p^2)+\frac{10}{9}]-\frac{g^2N_c}{16\pi^2}(\frac{4}{3}-2\xi+\frac{\xi^2}{4})
\end{array}\right)
\nonumber\\
&&\quad\quad+O(g^2)O_{E.O.M.}+O(g^2)O_{G.V.}+O(g^4)
\eea
\end{widetext}
where $O_{E.O.M.}$ and $O_{G.V.}$ label operators proportional to the equation of motion, and those operators which are gauge variant  (including the ghost operators) respectively. It is also the 1-loop matching coefficients that convert the renormalized EMT in the RI/MOM scheme to the $\overline{\textrm{MS}}$ scheme, when the condition $p^2=\mu_R^2$ is applied.

 We have computed the 1-loop correction for all those terms of $\overline{{\mathcal T}}^{\{\mu\nu\}}_{G,(0)}$ and have confirmed they are in good agreement with Ref.~\cite{Collins:1994ee} which is a good reference for details not presented here.

Combining with $R^{lat}$, we can get the renormalization matrix in the $\overline{\textrm{MS}}$ scheme for the EMT on the lattice,
\begin{widetext}
\bea
\left(\begin{array}{c}
\overline{{\mathcal T}}^{\mu\nu,\overline{\textrm{MS}}}_{Q}\\
\overline{{\mathcal T}}^{\mu\nu,\overline{\textrm{MS}}}_{G}
\end{array}\right)&=&
\left(\begin{array}{cc}
1-\frac{g^2C_F}{16\pi^2}[\frac{8}{3}\textrm{log}(a^2\mu^2)+\frac{40}{9}+B_{QQ}]&
0+\frac{g^2N_f}{16\pi^2}[\frac{2}{3}\textrm{log}(a^2\mu^2)+\frac{4}{9}+B_{GQ}] \\
0+\frac{g^2C_F}{16\pi^2}[\frac{8}{3}\textrm{log}(a^2\mu^2)+\frac{22}{9}+B_{QG}] &
1-\frac{g^2N_f}{16\pi^2}[\frac{2}{3}\textrm{log}(a^2\mu^2)+\frac{10}{9}+B^f_{GG}]\\
&-\frac{g^2N_c}{16\pi^2}(\frac{4}{3}+B_{GG})
\end{array}\right)
\left(\begin{array}{c}
\overline{{\mathcal T}}^{\mu\nu,lat}_{Q}\\
\overline{{\mathcal T}}^{\mu\nu,lat}_{G}
\end{array}\right)+O(g^4)
\nonumber\\
\eea
\end{widetext}
where $B_{QQ}$, $B_{GG}$ and $B^f_{GG}$ are the finite pieces in the 1-loop corrections of the quark and gluon EMT operator, $B_{QG}$ and $B_{GQ}$ are the finite peices in the 1-loop mixing between those two operators.

They can be obtained numerically and their values in different cases (the quark and gluon actions) are listed in Table~\ref{tab:mix}. The values with the superscript $T$ are the finite pieces in the traceless diagonal part of EMT and those without it are in the off-diagonal part. In the table, $B_{GG}=B_{GG}^v+B_{GG}^{v.t.}-B_G$ where the three terms are the contributions from the gluon EMT operator vertex without the tadpole term in Eq.~(\ref{eq:g4}), that tadpole term in the vertex, and the gluon self-energy contribution. The contribution of the tadpole term in the vertex is estimated following the strategy in Sec. \ref{sec:feynman_rules}. Note that the mixings from $O_{E.O.M.}$ and $O_{G.V.}$ are dropped here. The mixing from the equation of motion operators given by $O_{E.O.M.}$ should be the same in both lattice and continuum perturbation theory, which we have confirmed numerically.  The mixing from $O_{G.V.}$ is not relevant since the matrix element of a gauge variant operator is zero in lattice simulations.

In Table~\ref{tab:mix}, we listed several combinations of quark and gluon actions and the combined finite piece in the $\overline{\textrm{MS}}$ scheme are listed in Table~\ref{tab:mix2}.
On the quark slide, we listed the Wilson fermion and Overlap fermion (and also the case with the chiral fermion action $D_c$ in $B^f_{GG}$ since the correction with $D_c$ is different from that with the overlap fermion). Besides the case of the simplest Wilson gluon action, we also listed the results with the Iwasaki gluon action. In the simulation we proposed, we use the 1-step HYP smearing on the gluon action used for the fermion operator (marked as Iwasaki$^{HYP}$), and 5-step HYP smearing for the gluon operator (marked as $O^{5HYP}_g$), so one more column is added for this case. Note that the gluon action in the gluon operator renormalization case is still the original gauge action without any HYP smearing.

Our $B_{QQ}$,  $B_{QG}$ and $B^f_{GG}$ with Wilson action are consistent with those in Ref. \cite{Capitani:1994qn} except $B_{GQ}$. We have confirmed that $B_{QQ}$ and $B^T_{QQ}$ with the overlap action are consistent with those in Ref.~\cite{Horsley:2005jk} if we use to the same $\rho$ in the overlap action and the same gluon action. Note that the value of $B_{QQ}$ and $B^T_{QQ}$ are much larger with the overlap action than those with the Wilson fermion, if the HYP smearing is not applied. The contribution of the tadpole term in the vertex (Eq.~(\ref{eq:q2})) can be canceled by that in the quark self-energy in the Wilson fermion case, but this cancellation is not valid in the overlap fermion case.

\begin{table*}[!ht]
\begin{center}
\caption{\label{tab:mix} The finite pieces $B_{QQ,GQ,QG,GG}$ and $B^f_{GG}$ in the mixing between the quark and gluon EM tensor in different cases. The values with the superscript $T$ are the finite pieces in the traceless diagonal part of EMT and those without it are in the off-diagonal part. The value of $B_{GG}$ are split to several terms and the sum of them are listed in the last column. The values are those in the feynman gauge since the gauge dependence should vanish in the final renormalization matrix under $\overline{\textrm{MS}}$ scheme. See the text for more details.}
\begin{tabular}{c|cccccc|cc}
\hline
&\multicolumn{3}{c}{$B_{QQ}$} &\multicolumn{3}{c|}{$B^T_{QQ}$} & $B_{GQ}$ & $B^T_{GQ}$\\
& Wilson & Iwasaki & Iwasaki$^{HYP}$ &Wilson & Iwasaki & Iwasaki$^{HYP}$ &  \\
\hline
Wilson &  -3.17  & -2.59  & -1.53  &  -\ 1.88  & -\ 2.10  & -1.39 &   0.21& -0.56 \\
overlap & -36.96 & -20.00 & -5.25 & -36.40 & -19.78 & -5.14&   0.21& -0.81\\
\hline
&\multicolumn{3}{c}{$B_{QG}$} &\multicolumn{3}{c|}{$B^T_{QG}$} & \multicolumn{2}{c}{$B^f_{GG}=B^f_{G}$}\\
& Wilson & Iwasaki &  Iwasaki$^{HYP}$ & Wilson & Iwasaki &  Iwasaki$^{HYP}$&\\
&&&+$O_g^{5HYP}$&&&+$O_g^{5HYP}$&\\
\hline
Wilson &  -\ 5.82  &  -2.16  & 3.65 &  -\ 9.89  &  -3.50  & 3.50 & \multicolumn{2}{c}{-2.17}  \\
overlap & -\ 4.91 & -1.58  &  3.79 & -\ 9.03 & -3.12  &  3.64& \multicolumn{2}{c}{-0.72}\\ 
$D_c$ &  & &  &  & &  &\multicolumn{2}{c}{\ 0.16}\\ 
\hline
&\multicolumn{3}{c}{$B_{GG}$} &\multicolumn{3}{c|}{$B^T_{GG}$}&\\
& Wilson & Iwasaki &   Iwasaki & Wilson & Iwasaki &   Iwasaki &\\
&&&+$O_g^{5HYP}$&&&+$O_g^{5HYP}$&\\
\hline
$B_{GG}^v$ &\ \ \ 0.57   &  \ 3.13   & \ 0.92 & \ \ \ 2.47   & \ \ \  4.94     &\ \ \ 1.25\\
$B_{GG}^{v.t.}$ & -35.47 & -0.3(14.4)  & -14.6(4.1)   & -28.9(21.8)  &  9.1(13.8) &  -10.4(3.1)\\
-$B_G$ &\ 19.49  &-6.56 &-6.56  &  19.49  & -6.56 & -6.56\\
total & -15.58 & -4.0(14.4) &-20.5(4.4) &-6.9(21.8) & 7.5(13.8) &-15.7(3.1)
\end{tabular}
\end{center}
\end{table*}

\blue{
\begin{table*}[!ht]
\begin{center}
\caption{\label{tab:mix2} The finite pieces in the $\overline{\textrm{MS}}$ scheme. The values with the superscript $T$ are the finite pieces in the trace-less diagonal part of EMT and that without it are those in the off-diagonal part. The values reflect the difference between the lattice bare matrix elements and that in the $\overline{\textrm{MS}}$ scheme}
\begin{tabular}{c|cccccc|cc}
\hline
&\multicolumn{3}{c}{$\frac{40}{9}+B_{QQ}$} &\multicolumn{3}{c|}{$\frac{40}{9}+B^T_{QQ}$} & $\frac{4}{9}+B_{GQ}$ & $\frac{4}{9}+B^T_{GQ}$\\
& Wilson & Iwasaki & Iwasaki$^{HYP}$ &Wilson & Iwasaki & Iwasaki$^{HYP}$ &  \\
\hline
wilson & \ \ 1.27  & \ \ 1.85  & \ 2.91  &  \ \  2.56  & \ \  2.34  & \  3.05 &   0.65& -0.12 \\
overlap & -32.52 & -15.56 & -0.81 & -31.96 & -15.34 & -0.70 &   0.65& -0.37\\
\hline
&\multicolumn{3}{c}{$\frac{22}{9}+B_{QG}$} &\multicolumn{3}{c|}{$\frac{22}{9}+B^T_{QG}$} & \multicolumn{2}{c}{$\frac{10}{9}+B^f_{GG}$}\\
& Wilson & Iwasaki &  Iwasaki$^{HYP}$ & Wilson & Iwasaki &  Iwasaki$^{HYP}$&\\
&&&+$O_g^{5HYP}$&&&+$O_g^{5HYP}$&\\
\hline
wilson &  -\ 3.38  &  -0.28  & 6.09 &  -\ 7.45  &  -1.06  & 5.94 & \multicolumn{2}{c}{-1.06}  \\
overlap & -\ 2.47 & -0.86  &  6.23 & -\ 6.59 & -0.68  &  6.08 & \multicolumn{2}{c}{\ 0.39}\\ 
$D_c$ &  & &  &  & &  &\multicolumn{2}{c}{\ 1.27}\\ 
\hline
&\multicolumn{3}{c}{$\frac{4}{3}+B_{GG}$} &\multicolumn{3}{c|}{$\frac{4}{3}+B^T_{GG}$}&\\
& Wilson & Iwasaki &   Iwasaki & Wilson & Iwasaki &   Iwasaki &\\
&&&+$O_g^{5HYP}$&&&+$O_g^{5HYP}$&\\
\hline
$B_{GG}$ & -14.25 & -2.7(14.4) &-19.2(4.4) &-5.6(21.8) & 8.8(13.8) &-14.3(3.1)
\end{tabular}
\end{center}
\end{table*}
}

\section{Summary and outlook}\label{sec:summary}

The numerical result of the mixing matrix that match the lattice bare quantities to those under $\overline{\textrm{MS}}$ at $\mu=1/a$ with $g^2=3$ (or $\beta$=2.0 equivalently) is given for the case of the chiral fermion $D_c$ EMT operator with 1-step HYP smeared Iwasaki gauge link and 5-steps HYP smeared gluon EMT opeartor,
\bea\label{eq:fin_off}
\left(\begin{array}{c}
\overline{{\mathcal T}}^{\overline{\textrm{MS}}}_{Q}\\
\overline{{\mathcal T}}^{\overline{\textrm{MS}}}_{G}
\end{array}\right)&=&
\left(\begin{array}{cc}
1.0202 &
0.0123 N_f  \\
0.1565 &
2.08(25)-0.0239 N_f
\end{array}\right)
\left(\begin{array}{c}
\overline{{\mathcal T}}^{lat}_{Q}\\
\overline{{\mathcal T}}^{lat}_{G}
\end{array}\right)\nonumber\\
&&+O(g^4),
\eea
for the off-diagonal part of $\overline{{\mathcal T}}^{\mu\nu}$
and 
\bea\label{eq:fin_tr}
\left(\begin{array}{c}
\overline{{\mathcal T}}^{\overline{\textrm{MS}}}_{Q}\\
\overline{{\mathcal T}}^{\overline{\textrm{MS}}}_{G}
\end{array}\right)&=&
\left(\begin{array}{cc}
1.0175 &
-0.0069 N_f  \\
0.1528 &
1.84(18)-0.0239 N_f
\end{array}\right)
\left(\begin{array}{c}
\overline{{\mathcal T}}^{lat}_{Q}\\
\overline{{\mathcal T}}^{lat}_{G}
\end{array}\right)\nonumber\\
&&+O(g^4),
\eea
for the traceless diagonal part with the uncertainties coming from the estimate of the tadpole contributions. The 1-loop correction of the gluon operator is large which indicates the convergence problem for the perturbative series. 

Since the major contribution in this large correction comes from the tadpole in the gluon vertex and self-energy corrections, the cactus improvement~\cite{Constantinou:2006hz} procedure which resums major tadpoles contributions in order to achieve better convergence properties in lattice perturbative theory, would be helpful here. We will turn to the non-perturbative renormalization with 
RI/MOM scheme with the conditions listed in Eq.~(\ref{eq:condition_off})-(\ref{eq:condition_tr}) in the future to better investigate the renormalization of the quark and gluon momentum fractions.

\vspace{3cm}

\section*{ACKNOWLEDGMENTS}

This work is supported in part by the U.S. DOE Grant No.\ DE-SC0013065, DE-FG02-93ER-40762 and DE-AC02-05CH11231.  \mbox{Y. Y.} also thanks the Institute of High Energy Physics, Chinese Academy of Science for its partial support and hospitality. This material is also based upon work supported by the U.S. Department of Energy, Office of Science, Office of Nuclear Physics, within the framework of the TMD Topical Collaboration.
\bibliographystyle{apsrev4-1}
\bibliography{reference.bib}

\end{document}